\documentclass[prl,aps,twocolumn,groupedaddress]{revtex4}
\usepackage{graphicx}
\begin{document}

\title{\bf Quantum Hall Transition near a Fermion Feshbach Resonance in a Rotating Trap}

\author{Kun Yang$^1$ and Hui Zhai$^{2,3,4}$}
\affiliation{$^1$NHMFL and Department of Physics, Florida State
University, Tallahassee, Florida 32306, USA}
\affiliation{$^2$ Center for Advanced Study, Tsinghua University, Beijing, China, 100084}
\affiliation{$^3$ Department of Physics, University of California,
Berkeley, California, 94720, USA} \affiliation{$^4$ Materials Sciences Division,
Lawrence Berkeley National Laboratory, Berkeley, California, 94720,
USA}
\date{\today}
\begin{abstract}

We consider two-species of fermions in a rotating trap that interact via an s-wave Feshbach resonance, at total Landau level filling factor two (or one for each species). We show that the system undergoes a quantum phase transition from a fermion integer quantum Hall state to a boson fractional quantum Hall state as the pairing interaction strength increases, with the transition occurring near the resonance. The effective field theory for the transition is shown to be that of a (emergent) massless relativistic bosonic field coupled to a Chern-Simons gauge field, with the coupling giving rise to semionic statistics to the emergent particles.

\end{abstract}


\maketitle

Realization of paired fermionic superfluids represents a major breakthrough in the field of cold atom physics\cite{Jin,Ketterle}. Recently much attention has focused on the crossover from weakly paired (or BCS) atomic fermion superfluid to strongly paired bosonic molecular superfluid (or a molecular BEC state) as the pairing interaction increases, by sweeping the system through an s-wave Feshbach resonance (FR)\cite{review}. While there is very interesting universal many-body physics at the FR\cite{universality}, the whole process is a crossover that does not cross any phase boundary.

Neutral superfluids are very sensitive to rotation of the trap in which they live, as physically the effect of rotation is similar to that of a magnetic field in a charged superconductor\cite{zeeman}. When the rotation frequency is low the superfluid responds by introducing vortices, and the vortices form a regular static lattice. Such vortex lattices have been observed both in the BEC and BCS regimes\cite{Ketterle}. As the rotation frequency increases the vortex density also increases, and the vortex lattice is expected to melt eventually, resulting in a liquid state that is no longer a superfluid\cite{Hc2}. In this case the quantum mechanical nature of Landau level (LL) wave functions of the constituent particles play an important role, and theoretical studies suggest that in two dimensions (2D) various quantum Hall (QH) states are likely to form\cite{QH-atom}. It is very interesting to investigate what happens to these quantum Hall states as the pairing interaction between fermions changes, and in particular if quantum phase transitions (QPTs) occur near the FR instead of a simple crossover in the case without rotation.

\begin{figure}[bp]
\begin{center}
\includegraphics[angle=0,scale=0.4]
{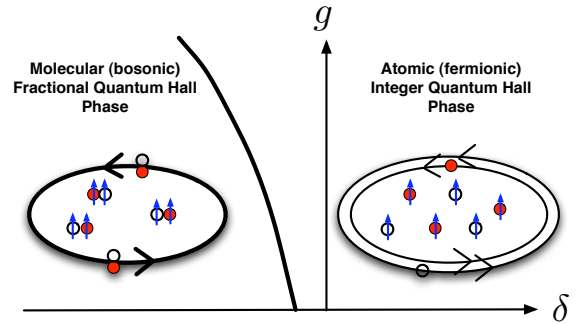}\caption{(Color on-line) Schematic phase diagram and illustration of two different quantum Hall phases. Solid (red) ball and open ball represent two different types of atoms, and short (blue) arrows represent a flux quantum attached to each atom. Each line with arrow represents a propagating edge mode. $\delta$ and $g$ represent detuning and atom-molecule coupling strength respectively (see text for details).\label{schematic}}
\end{center}
\end{figure}

In this paper we study the simplest realization of this situation, and show that QPT does occur between two very simple quantum Hall states at a specific fermion density or LL filling factor, namely $\nu_{\uparrow,\downarrow}=1$ for each of the fermion species (represented by ``spin"-1/2 indices $\uparrow$ and $\downarrow$ respectively). In the weak pairing regime where the paring interaction is much weaker compared to the LL spacing, the fermionic atoms form an integer QH (IQH) state. In the strong pairing regime the constituent particles become bosonic molecules; the number of molecules is equal to the number of each fermion species, but as its mass is doubled, the LL degeneracy is also doubled. Hence, the bosonic molecules will form a fractional QH (FQH) state of the Laughlin type with boson filling factor $\nu_b=1/2$\cite{QH-atom}. Although they have the same Hall conductance $\sigma_{xy}=2(e^*)^2/\hbar={1\over 2}[(2e^*)^2/\hbar]$ where $e^*$ is the fermion ``charge" and $2e^*$ is the boson ``charge"\cite{note}, they have different topological structures (as manifested by difference in quasiparticle quantum numbers and edge states), and must be separated by a phase boundary. We show that the effective field theory that properly describes this QPT is that of a massless relativistic bosonic field coupled to a Chern-Simons (CS) gauge field, with the coefficient of the CS term corresponding to semionic statistics of the particles. While these emergent particles carry non-trivial statistics, they are ``charge" neutral and do not contribute to the Hall conductance $\sigma_{xy}$, consistent with the fact that $\sigma_{xy}$ does not change across the phase boundary. Thus the emergent particles at the QPT carry quantum numbers {\em different} from the quasiparticles of both phases, a situation similar to that of the deconfined quantum critical points studied recently in other contexts\cite{senthil}.

We start by considering two limiting cases: (i) non-interacting limit, and (ii) the strong pairing limit where two fermions form a tightly bound bosonic molecule\cite{haldane}. For non-interacting fermions at $\nu_\uparrow=\nu_\downarrow=1$, both fermion species fully occupy the lowest LL and form an IQH state. The system supports two species of (gapped) {\em fermionic} quasiparticles with charge 1 (in unit of fermion charge $e^*$) and spin-1/2, and there are two branches of gapless chiral edge modes. In the strong pairing (BEC) limit, the bosons carry charge 2, and are at filling factor $\nu_b=1/2$; they form a bosonic Laughlin state. In this case the system supports a single species of Laughlin quasiparticles with charge ${1\over 2}\cdot 2=1$, spin zero and statistics angle $\pi/2$, or they are spinless {\em semions} with charge 1. Also there is a single branch of gapless chiral edge mode. These are thus two different topological phases that must be separated by a QPT; see Fig. 1 for an illustration.

The effective field theories for such simple QH states can be constructed in the following manner\cite{zhk}. We view each fermionic atom as a composite boson that carries a single flux quantum, which can be implemented by coupling the bosons to CS gauge fields (one CS field for each atom species). At mean field level the flux carried by the bosons is smeared out and cancels the external magnetic field (generated by rotation), and the bosons see zero field in average. The fermion IQH state corresponds to the phase in which both species of bosons are Bose condensed (thus there are two condensates). On the other hand in the strong pairing limit two charge-1 bosons form a charge-2 bound state corresponding to the molecules; these charge-2 bosons each carry two flux quanta, and the bosonic FQH state corresponds to the phase in which only this charge-2 boson condenses and there is no atom condensate in the system. As we show below, this field theory can also describe the QPT between these two phases.

The Euclidean Lagrangian density of the theory is
\begin{eqnarray}
L&=&L_\uparrow+L_\downarrow+L_{\text{m}}+L_{\text{cs}}
+g(\overline{\phi}\psi_\uparrow\psi_\downarrow+\phi\overline{\psi}_\uparrow\overline{\psi}_\downarrow),\label{L}\\
L_\sigma&=&\overline{\psi}_\sigma(\partial_\tau-a_0^\sigma)\psi_\sigma-\mu_\sigma|\psi_\sigma|^2\nonumber\\
&+&{1\over 2m_\sigma}|(-i \nabla-{\bf A}-{\bf a}^\sigma)\psi_\sigma|^2+\cdots,\label{La}\\
L_{\text m}&=&\overline{\phi}(\partial_\tau-a_0^\uparrow-a_0^\downarrow)\phi-(\mu_\uparrow+\mu_\downarrow-\delta)|\phi|^2\nonumber\\
&+&{1\over 2M}|(-i\nabla-2{\bf A}-{\bf a}^\uparrow-{\bf a}^\downarrow)\phi|^2+\cdots,\label{Lm}\\
L_{\text{cs}}&=&L^\uparrow_{\text{cs}}+L^\downarrow_{\text{cs}}={1\over 4\pi}{\pi\over \theta}\epsilon^{\mu\nu\lambda}\left[a^\uparrow_\mu\partial_\nu a^\uparrow_\lambda+a^\downarrow_\mu\partial_\nu a^\downarrow_\lambda\right].
\label{Lcs}
\end{eqnarray}
Here $\sigma=\uparrow,\downarrow$ is the atom ``spin" label, $\psi_\sigma$ and $\phi$ are boson fields representing the atoms and molecules respectively, ${\bf A}$ is the vector potential for the (static) external magnetic field (here it is due to rotation: ${\bf A}= m\Omega \hat{z}\times \bf{r}$ with $\Omega$ representing the rotational frequency), while ${\bf a}^\sigma$ is the CS gauge field that attaches flux to $\sigma$ particles. $\theta$ is the statistics angle after flux attachment of the $\sigma$ particles; in the present case $\theta=\pi$ corresponding to the fermionic statistics for the atoms. $\delta$ is the detuning; positive $\delta$ favors atoms while negative $\delta$ favors molecules, and FR corresponds to $\delta=0$. The molecule mass $M=m_\uparrow+m_\downarrow$, and the $g$ term in Eq. (\ref{L}) describes processes in which a pair of atoms turn into a molecule and vice versa\cite{g}. Generic density-density interactions among the particles are kept implicit and represented by $\cdots$.

The field theory (\ref{L}) has two independent U(1) symmetries corresponding to two separately conserved charges:
\begin{equation}
\label{conserved}
N_\sigma=\int{d^2{\bf r}}(|\psi_\sigma({\bf r})|^2+|\phi({\bf r})|^2).
\end{equation}
The CS term puts a constraint on the ${\bf a^\sigma}$ fields and conserved charge densities:
$
\nabla\times  {\bf a}^\sigma=2\theta(|\psi_\sigma|^2+|\phi|^2).
$
At $\nu_\sigma=1$, ${\bf A}$ and ${\bf a}^\sigma$ cancel each other in average:
$\nabla\times (\langle{\bf a}^\sigma\rangle+{\bf A})=0$, thus all
bosons see zero net flux.

We first consider a simpler version of the transition, in which the ${\bf a^\sigma}$ fields do not fluctuate and we can simply set all gauge fields to zero in Eq. (\ref{L}). In this case Eq. (\ref{L}) describes two species of {\em bosonic} atoms interact through a FR. In the weak pairing phase both U(1) symmetries in Eq. (\ref{conserved}) are broken because both types of atoms condense, and all three bosons fields acquire non-zero expectation values\cite{note1}; this is the atomic BEC (ABEC) phase. On the other hand in the strong pairing phase only a single U(1) symmetry that corresponds to conservation of $N_\uparrow+N_\downarrow$ is broken, and only the molecule field $\phi$ acquires a non-zero expectation value; this is the molecular BEC (MBEC) phase. Since the $\phi$ field is ordered in both phases, we can further simplify the problem by replacing $\phi$ with $\langle\phi\rangle$, and neglect its fluctuations (its Goldstone mode only has gradient couplings with remaining fields and are irrelevant), to obtain a reduced Lagrangian
\begin{equation}
\label{L'}
\tilde{L}=L_\uparrow[{\bf a^\sigma,A}=0]+L_\downarrow[{\bf a^\sigma,A}=0]+h(\psi_\uparrow\psi_\downarrow+\overline{\psi}_\uparrow\overline{\psi}_\downarrow),
\end{equation}
where $h=g\langle\phi\rangle$ (assumed to be real and positive without loss of generality). To determine the critical point, we need to diagonalize the quadratic terms (that do not involve derivatives) in Eq. (\ref{L'}), and this can be done easily for the symmetric (or balanced) case for $\uparrow$ and $\downarrow$ atoms with $\mu_\uparrow=\mu_\downarrow=\mu$ and $m_\uparrow=m_\downarrow=m$ by introducing
\begin{equation}
\label{mix}
\psi_+=(\psi_\uparrow+\overline{\psi}_\downarrow)/\sqrt{2}; \hskip 0.5cm \psi_-=(\psi_\uparrow-\overline{\psi}_\downarrow)/\sqrt{2}.
\end{equation}
In terms of $\psi_+$ and $\psi_-$, $\tilde{L}$ takes the form (after neglecting total derivatives)
\begin{eqnarray}
\tilde{L}&=&\overline{\psi}_+\partial_\tau\psi_-+\overline{\psi}_-\partial_\tau\psi_+ -(\mu+h)|\psi_-|^2+(h-\mu)|\psi_+|^2\nonumber\\
&+&{1\over 2m}\left[|\nabla\psi_-|^2+|\nabla\psi_+|^2\right]+\cdots.
\end{eqnarray}

We note that the presence of a molecular condensate suggests that the molecule chemical potential $2\mu-\delta\approx 0$, i.e. $\mu\approx \delta/2 < 0$ on the MBEC side (of course the actual value of the chemical potential receives corrections from density-density interaction and is not universal). As we approach resonance,
$\psi_-$ will reach criticality (or becomes massless) when $\mu+h=0$, (i.e. when $\delta\approx -2h$, on the molecule side of the resonance) at which the MBEC will be unstable and undergo phase transition to an ABEC. Note that at the critical point $\psi_+$ remains massive ($h-\mu=2h > 0$); it is thus natural to integrate out $\psi_+$ to obtain an effective theory in terms of the critical field $\psi_-$:
\begin{eqnarray}
\label{L-}
L_{\text {eff}}[\psi_-]&=&{|\partial_\tau\psi_-|^2\over h-\mu}+{|\nabla\psi_-|^2\over 2m}-(\mu+h)|\psi_-|^2\nonumber\\
&+&U|\psi_-|^4+\cdots,
\end{eqnarray}
in which we included the quartic interaction that is relevant at the critical point. The effective field theory (\ref{L-}) is the relativistic $\phi^4$ theory with U(1) symmetry at $2+1$ dimensions, with (mean-field) critical point at $\mu+h=0$; the critical behavior is that of the 3D XY model\cite{kuklov}.

For the more generic case where the two atom species are imbalanced, {\em e.g.} due to a chemical potential difference: $\mu_\uparrow\ne\mu_\downarrow$, Eq. (\ref{L'}) needs to be diagonalized using a more general transformation of the form
\begin{equation}
\label{}
\psi_+=u\psi_\uparrow+v\overline{\psi}_\downarrow; \hskip 0.5cm \psi_-=v\psi_\uparrow-u\overline{\psi}_\downarrow
\end{equation}
with $u,v$ real and $u^2+v^2=1$. In this case in addition to the terms in (\ref{L-}), there is a term with {\em first} derivative in $\tau$ of the form $(u^2-v^2)\overline{\psi}_-\partial_\tau\psi_-$ in the effective field theory $L_{\text{eff}}[\psi_-]$. This term is more relevant than the $|\partial_\tau\psi_-|^2$ term, and dominates the dynamics. In particular it changes the dynamical exponent $z$ from 1 to 2, and ruins the Lorentz invariance of (\ref{L-}); the universality class of the MBEC to ABEC transition becomes that of the dilute Bose gas transition\cite{sachdevbook} in this case.

We now return to the problem of QH transition, in which the fluctuations of CS gauge fields are crucial. For simplicity we will focus on the balanced case with $\mu_\uparrow=\mu_\downarrow=\mu$ and $m_\uparrow=m_\downarrow=m$. Similar to Eq. (\ref{mix}), we introduce new combinations of CS gauge fields:
\begin{equation}
\label{csmix}
a_\mu^+=(a_\mu^\uparrow+a_\mu^\downarrow)/2, \hskip 0.5cm a_\mu^-=(a_\mu^\uparrow-a_\mu^\downarrow)/2,
\end{equation}
in terms of which Eqs. (\ref{La},\ref{Lm},\ref{Lcs}) take the form
\begin{eqnarray}
\label{Lup}
L_\uparrow&=&\overline{\psi}_\uparrow(\partial_\tau-a_0^+-a_0^-)\psi_\uparrow-\mu|\psi_\uparrow|^2\nonumber\\
&+&{1\over 2m}|(-i\nabla-{\bf A}-{\bf a}^+-{\bf a}^-)\psi_\uparrow|^2+\cdots,\\
\label{Ldown}
L_\downarrow&=&\overline{\psi}_\downarrow(\partial_\tau-a_0^++a_0^-)\psi_\downarrow-\mu|\psi_\downarrow|^2\nonumber\\
&+&{1\over 2m}|(-i\nabla-{\bf A}-{\bf a}^++{\bf a}^-)\psi_\downarrow|^2+\cdots,\\
L_{\text{m}}&=&\overline{\phi}(\partial_\tau-2a_0^+)\phi-(2\mu-\delta)|\phi|^2\nonumber\\
&+&{1\over 2M}|(-i\nabla-2{\bf A}-2{\bf a}^+)\phi|^2+\cdots,\\
L_{\text{cs}}&=&{1\over 4\pi}{\pi\over \theta'}\epsilon^{\mu\nu\lambda}\left[a^+_\mu\partial_\nu a^+_\lambda+a^-_\mu\partial_\nu a^-_\lambda\right].
\end{eqnarray}
Here $\theta'=\theta/2=\pi/2$; this suggests $a^+$ and $a^-$ each attaches one {\em half} flux quantum to a unit charge coupled to them. Physically one can understand this as $a^+$ attaches half flux quantum to the {\em charge} of the atom, and $a^-$ attaches half flux quantum to the {\em spin} of the atom, so each atom carries one flux quantum in total.

To proceed we make the following observations. (i) The molecular field $\phi$ couples to $a^+$ with charge 2, but does not couple to $a^-$. Physically this is because the molecule carries twice the atom charge, but does not carry spin. (ii) The molecular field $\phi$ acquires a nonzero expectation value in both phases and at the QPT; it thus gives a Higgs mass term to the fluctuation of ${\bf a}^+$, $\Delta{\bf a}^+={\bf a}^+-\langle{\bf a}^+\rangle={\bf a}^++{\bf A}$, of the form $(2|\langle\phi\rangle|^2/M)|\Delta{\bf a}^+|^2$. Physically this reflects the fact that we are in a QH state throughout the phase diagram, thus the system is incompressible and charge fluctuations are suppressed. We can thus replace $\phi$ by $\langle\phi\rangle$, and safely integrate out the massive field $\Delta {\bf a}^+$, after which we obtain
\begin{eqnarray}
L'&=&L'_\uparrow+L'_\downarrow+L^-_{cs}+h(\psi_\uparrow\psi_\downarrow+\overline{\psi}_\uparrow\overline{\psi}_\downarrow),\\
L'_\uparrow&=&\overline{\psi}_\uparrow(\partial_\tau-a_0^-)\psi_\uparrow-\mu|\psi_\uparrow|^2\nonumber\\
&+&{1\over 2m}|(-i\nabla-{\bf a}^-)\psi_\uparrow|^2+\cdots,\\
L'_\downarrow&=&\overline{\psi}_\downarrow(\partial_\tau+a_0^-)\psi_\downarrow-\mu|\psi_\downarrow|^2
\nonumber\\
&+&{1\over 2m}|(-i\nabla+{\bf a}^-)\psi_\downarrow|^2+\cdots,\\
L^-_{\text{cs}}[a^-]&=&{1\over 4\pi}{\pi\over \theta'}\epsilon^{\mu\nu\lambda}a^-_\mu\partial_\nu a^-_\lambda={1\over 2\pi}\epsilon^{\mu\nu\lambda}a^-_\mu\partial_\nu a^-_\lambda.
\end{eqnarray}
Now performing the transformation (\ref{mix}), we obtain
\begin{eqnarray}
&L'&=\overline{\psi}_+(\partial_\tau-a^-_0)\psi_-+\overline{\psi}_-(\partial_\tau-a^-_0)\psi_+ \nonumber\\
&-&(\mu+h)|\psi_-|^2+(h-\mu)|\psi_+|^2+(1/4\theta')\epsilon^{\mu\nu\lambda}a^-_\mu\partial_\nu a^-_\lambda\nonumber\\
&+&{1\over 2m}\left[|(\nabla+i{\bf a}^-)\psi_-|^2+|(\nabla+i{\bf a}^-)\psi_+|^2\right]+\cdots.
\end{eqnarray}
We can again integrate out the massive field $\psi_+$ to arrive at the effective field theory for the QH transition, in terms of the critical field $\psi_-$ and the CS field $a^-$:
\begin{eqnarray}
\label{Leff}
&&L_{\text{eff}}[\psi_-, a^-]={|(\partial_\tau-a_0^-)\psi_-|^2\over h-\mu}+{|(\nabla-i{\bf a}^-)\psi_-|^2\over 2m}\nonumber\\
&-&(\mu+h)|\psi_-|^2+U|\psi_-|^4+{1\over 4\theta'}\epsilon^{\mu\nu\lambda}a^-_\mu\partial_\nu a^-_\lambda+\cdots.
\end{eqnarray}
This is the central result of this paper.

The molecule FQH to atom IQH transition is driven by the condensation of $\psi_-$, occurring at $\delta_c\approx -2h$ as in the previous case. Fig. 1 is a schematic phase diagram of the system.
Notice that $\psi_-$ describes an emergent particle whose quantum numbers differ from those of the quasiparticles of both phases: It has spin-1/2, statistics angle $\theta'=\pi/2$ (or it obeys semionic statistics), yet carries zero atom charge as it is a half-half mixture between particle and hole (see Eq. (\ref{mix})), and does not couple to external gauge field that couples to the charge. This phenomenon is similar to what happens at deconfined critical points discussed elsewhere\cite{senthil}.

The appearance of a new condensate $\langle\psi_-\rangle$ adds one additional edge mode to the system, thus there are two edge modes in the atom IQH phase. It also modifies the quasiparticle properties in the following manner. In the molecule FQH phase, the quasiparticles are vortices of the charge-2 molecule field $\phi$, which also carries {\em one half} flux quantum in ${\bf a}^+$. But this causes frustration when there is a $\psi_-$ condensate, which implies the the presence of condensates of $\psi_\uparrow$ and $\psi_\downarrow$. This is because $\psi_\uparrow$ and $\psi_\downarrow$ couple to ${\bf a}^+$ as charge-1 fields, thus incompatible with a half flux quantum. To resolve this conflict the vortex of $\phi$ must also bind a half flux quantum in ${\bf a}^-$ with either $+$ or $-$ sign, such that one of the $\psi$ fields sees zero total flux while the other sees one unit flux quantum [see Eqs. (\ref{Lup},\ref{Ldown})]. This composite object is a vortex in one of the $\psi$ fields that binds one flux quantum in the corresponding CS field, and becomes the new quasiparticle in the atom IQH phase. It is easy to see that it carries charge 1, spin-1/2, and obey fermionic statistics, which are precisely the properties of the quasiparticles of the atom IQH state. We note that theoretical proposals of measuring quasiparticle statistics are availble\cite{zoller}, which allow for experimental distinction between these two phases.

The critical behavior of the effective theory (\ref{Leff}) was first studied by Wen and Wu\cite{wenwu}, as a theory of quantum Hall-insulator transition driven by a periodic potential in electronic quantum Hall systems. In those systems such a transition has never been realized; thus the trapped cold atom system with rotation discussed here may well become an alternative experimental realization of the theory (\ref{Leff}). In a large-$N$ limit Wen and Wu found a 2nd-order transition with $\theta'$ dependent critical behavior. On the other hand Pryadko and Zhang\cite{pz} presented arguments against $\theta'$-dependent critical point by showing that the transition is driven 1st-order by gauge fluctuations in certain parameter regimes of (\ref{Leff}). Thus the critical behavior of (\ref{Leff}) remains an open problem. Experiments in cold atom systems can shed light on this important issue. If the QPT is 2nd order, then the spin gap $\Delta_s$  vanishes as one approaches the QPT: $\Delta_s\sim |\delta-\delta_c|^{z\nu}$, where $\nu$ is correlation length exponent and the dynamical exponent $z=1$ due to Lorentz invariance. One can measure the spin gap using radio frequency (RF) spectroscopy, if the two species of the atoms are two different hyperfine states of the same atom, and a transition between the two can be induced by RF radiation. The exponent $\nu$ extracted from data can be compared with theory\cite{wenwu}.

We thank Kavli Institute for Theoretical Physics (KITP, which was
supported in part by the NSF under Grant No.
PHY05-51164), and Kavli Institute for Theoretical Physics China
(KITPC) for hospitality. This
work was supported by NSF grant No.
DMR-0225698 (KY), and by NSFC under grant No. 10547002 (HZ).


\begin{references}
\bibitem{zeeman} Rotation only has orbital effect and does not lead to any Zeeman splitting between different ``spin" species.
\bibitem{note} Not to be confused with electric charge; both the fermions and bosons are assumed to be neutral here.
\bibitem{g} In microscopic calculation $g$ need to be renormalized in order to eliminate the short-range (high-energy) divergency. Here as we are only interested in low-energy behavior of the system, the effective field theory will not be affected by the details of renormalization scheme.
\bibitem{note1} Due to the $g$ term in Eq. (\ref{L}), $\langle\psi_\uparrow\psi_\downarrow\rangle\ne 0$ forces $\langle\phi\rangle\ne 0$ and {\em locks the phase} of $\phi$. However the reverse is {\em not} true; $\langle\phi\rangle\ne 0$ does not guarantee the same for $\psi$'s.
\end{references}
\end{document}